# DistB-SDoIndustry: Enhancing Security in Industry 4.0 Services based on Distributed Blockchain through Software Defined Networking-IoT Enabled Architecture

Anichur Rahman[1]
Department of Computer Science and Engineering
Mawlana Bhashani Science and Technology University
Tangail, Bangladesh
National Institute of Textile Engineering and Research
(NITER), Savar, Dhaka, Bangladesh

Umme Sara[2]
Department of Computer Science and Engineering
National Institute of Textile Engineering and Research
(NITER), Dhaka, Bangladesh

Dipanjali Kundu[3]
Department of Computer Science and Engineering
National Institute of Textile Engineering and Research
(NITER), Savar, Dhaka, Bangladesh

Saiful Islam[4]
Department of Computer Science and Engineering
International Islamic University Chittagong
Chittagong, Bangladesh

Md. Jahidul Islam[5]
Department of Computer Science and Engineering
Green University of Bangladesh
Dhaka, Bangladesh

Mahedi Hasan[6]
Department of Computer Science and Engineering
National Institute of Textile Engineering and Research
(NITER), Dhaka, Bangladesh

Ziaur Rahman[7]
Department of Information and Communication Technology
Mawlana Bhashani Science and Technology University
Tangail, Bangladesh

Mostofa Kamal Nasir[8]
Department of Computer Science and Engineering
Mawlana Bhashani Science and Technology University
Tangail, Bangladesh

*Abstract*—The concept of Industry 4.0 is a newly emerging focus of research throughout the world. However, it has lots of challenges to control data, and it can be addressed with various technologies like Internet of Things (IoT), Big Data, Artificial Intelligence (AI), Software Defined Networking (SDN), and Blockchain (BC) for managing data securely. Further, the complexity of sensors, appliances, sensor networks connecting to the internet and the model of Industry 4.0 has created the challenge of designing systems, infrastructure and smart applications capable of continuously analyzing the data produced. Regarding these, the authors present a distributed Blockchain-based security to industry 4.0 applications with SDN-IoT enabled environment. Where the Blockchain can be capable of leading the robust, privacy and confidentiality to our desired system. In addition, the SDN-IoT incorporates the different services of industry 4.0 with more security as well as flexibility. Furthermore, the authors offer an excellent combination among the technologies like IoT, SDN and Blockchain to improve the security and privacy of Industry 4.0 services properly. Finally , the authors evaluate performance and security in a variety of ways in the presented architecture.

## I. INTRODUCTION

Mostly, the monitoring and control mechanism in industries that manages specifics of commodity production, inventory data, knowledge of employees working in the supply chain, is typically time-intensive, costly, and sluggish. Industry 4.0, often referred as the Industrial Internet of Things (IIoT), is a modern step of the Industrial Revolution, focused exten- sively on interconnectedness, robotics, artificial intelligence, and real-time data. In addition, the smart industry has four intelligent features. Firstly, sensors that make decisions to alter behaviour based on environmental changes. Secondly, it has internet connectivity and real-time access. Then, IR 4.0 is strongly compatible with robot vision systems and AI techniques. The last one is Virtual Reality (VR) strategies that enable human-machine interaction in IR-4.0 [1]. However, the IT protection issue is the most daunting part of the adoption of industry 4.0 platforms or strategies. Further, the key issues in Industry 4.0 are the absence of granularity of knowledge and real-time tracking [2].

On the contrary, the SDN [3], [4] is an evolving technology that enables the implementation of a protected Industry 4.0 manufacturing environment [5], because low-level computing functions in SDN are far more effective, authentication functions are embedded in the network rather than being centralized in individual network components. Hence the key aim of introducing SDN is to minimize public response time and continuous availability. These technologies do have specific types of implementations in various smart technical fields, such as smart buildings, grids, healthcare, industries, and many more. Furthermore, research into security mechanisms is critical for next-generation IoT and the creation of advanced confidentiality defence schemes to tackle numerous attacks on IoT networks. To deliver influential functionality such as continuing anonymity, verification, and robustness; Blockchain technology is a secure solution [6]. In addition to this, Blockchain accounts for new technology and developments of the future. The invention of the Blockchain is a radical change of the conventional societal structure and style of service. Due to the popularity of Ethereum and Bitcoin cryptocurrency, Blockchain has gone authoritative in the field of network security. It creates immutable data structures that cannot be hacked, changed. The process of new block data appendage is performed by some proof of work and acknowledging the current blocks of data appended already in the decentralized public ledger of the Blockchain [7].

Several researchers have investigated the importance of various innovations for developing the smart industry or IR 4.0 [8], [9]. Nevertheless, such innovations cannot address the current problems of Industry 4.0 alone, such as stability, data protection, etc. On the other hand, utilizing all the new technology together, such as SDN, IoT, Blockchain, deep learning, etc. techniques would contribute to complicated structures. However, the IoT, Blockchain, SDN innovations are integrated to reach a higher degree of operating performance, profitability, transparency, protection, and privacy in Industry 4.0. Within IR 4.0 major focus is paid to SDN, Blockchain and IoT, etc.

After analyzing the above discussion, this study proposed a model based on distributed Blockchain through SDN-IoT enabled architecture to ensure adequate security which is the primary concern in industry 4.0. Moreover, the authors focus on the significant concerning issues of Industry 4.0 applications like security, privacy, confidentiality efficiently.

The main contributions of the paper are following:
- This study proposes a framework "DistB-SDoIndustry" focused on both SDN-IoT and Blockchain technologies to control more securely in Industry 4.0 services.
- We also address Blockchain technology for data validation, verification, broadcasting, and so on. Additionally, it is capable of providing data with a confidential route to enter the desired cloud locations efficiently.

The remaining sections of the paper are structured as follows. The authors present related works in Section II. Section III highlight the proposed "DistB-SDoIndustry" architecture for industry 4.0 security management. After that result analysis and discussions are also presented in Section IV. Finally, this work concludes with the significance and future ramifications in Section V.

## II. MOTIVATIONAL BACKGROUND AND LITERATURE REVIEWS

### A. Background Study

In this section, the authors discuss the IoT, SDN and background knowledge of Blockchain technology with Industry 4.0 applications briefly.

*1) IoT with SDN technologies:* The IoT includes various types of information advancing devices like routers, different environment detector which can perceive data from surroundings and pass the data to the next level of digital system [10]. IoT can choke the data collected from the natural world but not securely and conveniently. To handle the information smartly, the compound structure of IoT, including Software Defined Network (SDN) is helpful. SDN controls the information through multiple central processors and makes the system pliant with the reward of programmability [11], [12]. It could have pertained as a secret agent between the data control layer and IoT by which information is collected.

*2) Overview of Blockchain Concept:* Blockchain [13] is a system of recording data in a way that makes it difficult or impossible to change, hack, or cheat the system. A Blockchain is a decentralized, distributed, and public digital which consists of blocks. In general, every block is connected between them and sets of timestamped transactions. The chain of blocks, or Blockchain, serves as a publically accessible digital ledger. In this technology node exchange data by creating a transaction and each transaction depends on the previous transaction, where one transaction outputs are connected in another transaction as inputs hence forming chain among them. The Blockchain representation is shown in Fig. 1. The first block is called a generic block, and the rest of the blocks create by participating nodes called miners to try to solve a cryptographic puzzle named Proof of Works. Hence, participating nodes create a trusted network over untrusted participants in the network. New transactions are verified by all participating nodes that exclude the requirement of the central dependency and propose a distributed management system. Each block holds the hash of its previous block which assures the integrity of the transaction; therefore, make sure no alteration of the block in the network. If one transaction is valid, then the transaction is continuously stored in the Blockchain network that can be reached by any node. All transactions in this network are signature using public-key cryptography so that the authenticity nature of Blockchain is fulfilled [14].

*3) Blockchain for Industry 4.0:* Industry 4.0 aims to assemble, study, disc the information of individuals and assure the activities in real-time. Today's world is having a lot of

676 | P a g e

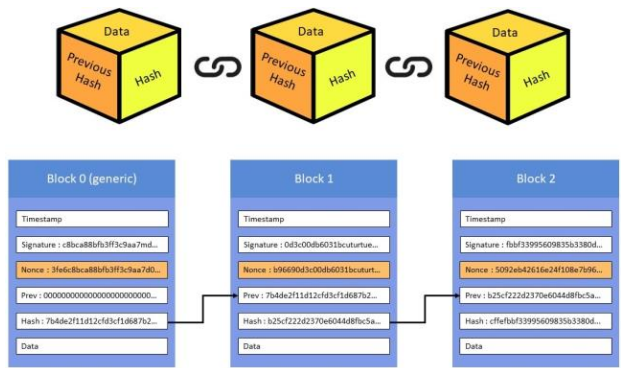

Fig. 1. Concept of Blockchain

industries, and each industry deals with a vast amount of products and customers. This information helps the industry to take the decision about manufacturing products in future. It is quite backbreaking to take care of the information of yield and clients and predicting the doings [15]. IoT with other relevant technologies extensively fires the 4th industrial revolution [16]. It can be considered as a bright set up of a factory and extraordinarily dynamic and automated production mesh. Moreover, it is to develop the manual manufacturing process into an elastic and self-coordinating production chan- nel [17]. In a real-time machine, it is quite risky to store and control data as a lot of actuators are collecting information from different places. Blockchain comes up at this point. As mentioned earlier Blockchain can secure data through its unique structure and processing power. It is not impossible to litigate a transaction of the product without any human resources using Blockchain.

### B. Literature Review

Several researchers have been proposed in recent years based on Blockchain and SDN with IoT technologies in various purposes. Throughout this portion, we are going to present similar studies based on IoT convergence with SDN and Blockchain for Industry 4.0 applications.

In [18] through this contribution, the enhanced framework for considering the smart industry can be demonstrated by using IoT. Moreover, the authors taken an energy-efficient approach into account for Industry 4.0 but not mentioned stability. On the similar research, an IoT network protection framework by using the core technologies provided by the SDN proposed by [19] such an advanced protection strategy including IoT system authentication and authorizing approved flows will help secure IoT networks from malicious IoT devices and attacks. However for Peer to Peer (P2P) con- nectivity between IoT systems and SDN controllers, [20] proposed design that utilizes public and private Blockchains, to remove Proof-of-Work (POW), in addition, the proposed model has used cluster configuration routing algorithm to maximize energy usage and improve protection. On the other hand, in [6] introduced a protection architecture and applied it as a Blockchain-based Platform-as-a-service (PaaS) paradigm to validate the data secrecy of authenticated users while applications are on the move and provide a robust solution to threat detection for usage in the IoT context. The proposed model for protection is efficient. From a virtualization viewpoint, it can be improved, maintaining certain protection features such as secrecy, transparency, etc. Again, the similar work [21] authors proposed "DistBlockBuilding" framework to handle the stable and efficient movement of data from one surface to another, also assessed the efficiency of a protected network based on IoT-SDN architecture. They suggested a cluster head selection algorithm; in addition, they included energy-saving and load balancing resources for SDN-IoT infrastructure in distributed Blockchain-based network. Further, [22] authors developed 5G network intrusion detection and mitigation technologies for the SDN/NFV cloud. Additionally, [23] focused Blockchain defi- nition that can be converged to an SDN based IoT framework to strengthen its protection aspects further.

In another research growth, in [24] discusses the effects of Blockchain for IoT also stresses how IoT can take advantage of the Blockchain's decentralized, arbitrary and transparent nature to improve agility in asset management. But they over- looked Blockchain's technological qualities. Therefore, the practicalities of applying the core characteristics have not been assessed. Moreover, Rane et al. [25] suggested Man4Ware architecture with the supplementary Blockchain infrastructure. Also, the creation of smart manufacturing software compatible with the Industry 4.0 vision with "Man4Ware" support will generate enough incentives for innovative and insightful apps. But the proposed architecture does not include more sophis- ticated functionality and help for technological innovations and Blockchain apps. To render Blockchain entirely available and customizable [17] reviewed the current research on the applicability of Blockchain in various IIoT-specific industries. In Industry 4.0 and IIoT, they looked at the different industrial application of Blockchain to include an abstract indicator of acceptance in practice; Then, they added that in order for Blockchain to be completely functional and scalable, industry-oriented work would also tackle several of these problems, including personal data security, block system scalability, the participating organization's data confidentiality and safety, Blockchain deployment and implementation costs, and policy regulations. Besides, [26] have defined IIoT technologies and have addressed the state-of-the-art protection vulnerabilities in Industry 4.0. Furthermore, Oztemel et al. [27] analyzed Indus-try 4.0, and associated innovations also [28] reviewed some of the cyber-security threats included in the advanced industrial production and Industry 4.0 path, further the relevant preven-tive measures currently adopted or under implementation. On the similar work to introduced an energy-efficient and QoS-aware parallel routing optimization algorithm for Software-Defined IIoT focused on healthcare systems, a scalable routing scheme for an outsized IIoT network has proposed [29] which is eight times faster than the existing programmes. Moreover, a quick parallel online routing optimization architecture is pro-posed for SDN-enabled smart healthcare networks supported IIoT.

In summary, existing researches have focused on many branches of Blockchain, IoT-SDN technologies. But only a few number of researchers have addressed Industry 4.0 applications. As the field is still new and, is facing many threats in security correspondence. That's why, we attempt to minimize challenges such as security threats in the Industry 4.0 applications.

## III. PROPOSED "DISTB-SDOINDUSTRY" FOR INDUSTRY 4.0 APPLICATIONS

From the above discussed sections we can conclude that Industry 4.0 is vulnerable to security issues.To manage a massive amount of data that is transferred need to be maintained efficiently and securly.In addition to this the privacy of data also need to ensured as the data transferred over the internet is highly susceptible to attackers.For managing the different applications of Industry 4.0, the authors propose a model "DistB-SDoIndustry" based on emerging technologies like SDN-IoT and Blockchain in security purposes, which shown in Fig. 2. We consider several steps for elaborating the proposed architecture, such as SDN-IoT enabled environment with perception, control and application layers, distributed Blockchain-based security in Industry 4.0 credentials, and Industry 4.0 services and security. In addition, the whole architecture is controlled by SDN and Blockchain technologies. Indeed, SDN helps to provide the programmability and data flexibility to the Industry 4.0 environment. Then, Blockchain is capable of handling the security, privacy as well as the confidentiality of the desire networks efficiently [30].

### A. SDN-IoT Enabled Environment

Basically, the IoT sensor is capable of sensing data for desire applications. The IoT provides these data-enabled devices like firewalls, routers, switches, and so on. On the other hand, SDN helps to ensure data security and flexibility in the IoT environment. Indeed, an SDN is organized by some distinct planes such as data, network, and application planes efficiently. In addition, discussion of these layers is given below.

*1) Perception Layer:* First of all, this layer is responsible for providing data to the Industry 4.0 management. It can collect huge data for future use in the goal system perfectly. However, the initial sources of data forwarding are treated as smart sensor, actuators, and core smart grid networks. Moreover, it can perform and monitor all data effectively. In SDN platform, it estimates the data path management regionally on this layer using compromises node-to-node in real-time standings. On the other hand, requests for more extended bandwidth or source may be difficult to meet the device requirement in real-time, as this is unpredictable in- formation subject to application and traffic Profile. Moreover, an SDN data plane manages the whole gathered data using a common gateway path; this gateway incorporates data as well as filtering this from external interruptions.

*2) Control Layer:* In the SDN paradigm, a most essential plane is the control path. This basically provides all benefits of the SDN platform like data control and data security also control the transmission of data from the edge layer to manage layer efficiently. Therefore, the workflow of the control layer in the SDN environment, several protocols like OpenFlow, OpenDayLight, and OpenStack used. However, it interfaces of two Application Program Interface (API), such as southbound and northbound APIs. Where the Southbound APIs features, provide connectivity with the switch fabric, virtualized network frameworks, or the consolidation of a decentralized network of computers. On the other hand, a northbound interface is an interface that allows the communi- cation with a higher-level feature of a specific component of a device. Nevertheless, the entire data path can be updated if the application layer demands a change based on the gathering of data, availability of contact connections and a distribution scheme.

*3) Applications Layer:* The application layer is the topmost tier of the SDN stages, and it involves data centre, servers, storage, analysis, processing, and statements. It efficiently processes all data. Further, SDN application layer collects all filtered data from SDN controllers applying suitable instructions of the control layer. After performing all operation of data in the data and control layer, the application layer consists of all filtered data suitably. Moreover, it helps to store these data in the cloud storage using a secure communication path. Then, Industry 4.0 entertains these secured data in various applications efficiently.

### B. Distributed Blockchain-based Security in Industry 4.0 Applications

There are many challenges in the growing Industry 4.0 applications; these challenges are cost challenge, structural, technology, security challenges, and so on. Moreover, Industry 4.0 applications are facing data integrity and data redundancy problems. In this paper, the author's primary concern is security challenges. Indeed, the authors also highlighted security improvement using the Blockchain approach for Industry 4.0 services management. In the presented model, the Blockchain approach provides the solution such a concerning issues like confidentiality, access control, authorization, and integrity in the Industry 4.0 applications. Moreover, a Blockchain is a chain based strategy, which provides the chain to connect every block to each other. Each block consists of data, timestream, hash data, and so on, as depicted in Fig. 3. In the Blockchain scenario, hash data is a unique component that can point one block to another. Every block of information is connected with one to another using the hash function properly. First of all, the genesis block creates by the system, then the rest of the block continues their activity based on the genesis block. If a new block is added to the Blockchain environment, it will need get the permission of all minor blocks. When it achieves the consent of at least 60-80 percent block, it will be added in the Blockchain network as a new member securely. This process is performed very confidentially and firmly. In contrast, there is no involvement of a third party or any intruder in this communication system. For the benefits of Blockchain, we have addressed this technology in the proposed method. In Industry 4.0 applications, all data would be safe by using Blockchain technology. Then, it provides different services like data validations, cleaning, broadcasting, as well as data capturing & query interfacing in the modern Industry 4.0 environment, as shown in Fig. 2. Furthermore, it can manage various security attacks and also able to perform a large number of operations in the Industry 4.0 applications.

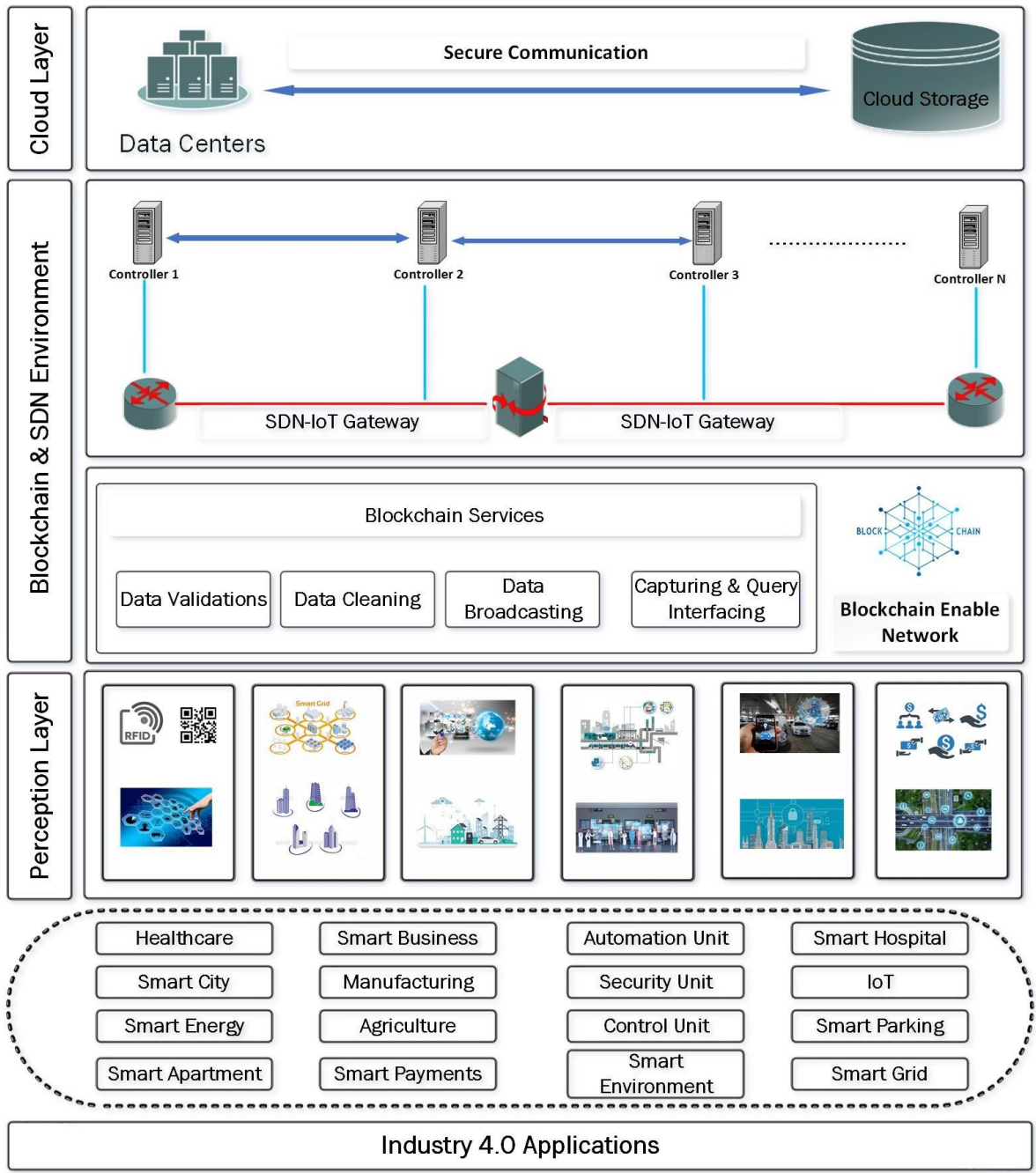

Fig. 2. Proposed Architecture of "DistB-SDoIndustry"

C. *Industry 4.0 Services and Securities*

Modern Industry 4.0 provides some benefits like efficiency in automation, the innovation of new products, costs consideration, revenues, etc. [31]. It can also offer intelligent components such as connectivity, automation, and optimization. Moreover, the IIoT consists of all sensors and machines in the Industry 4.0 platform.

In addition, automation means the digitalization of all services for Industry. After that Industry 4.0 includes some leading technologies like AI, whose main contributor is Big

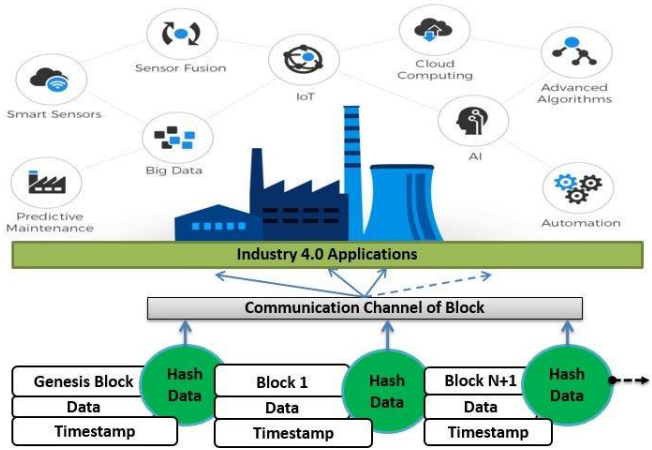

Fig. 3. Blockchain-based Security Approach in Industry 4.0

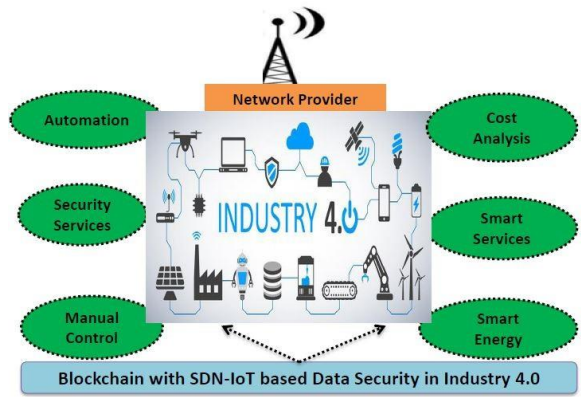

Fig. 4. Industry 4.0 System and Services

Data and Analytics, which utilizes machine learning and AI techniques efficiently. Based on these services of Industry 4.0, the authors mention different services such as security services, smart services, as well as smart energy, as shown in Fig. 4. Indeed, the authors also address various types of services such as smart cities, smart apartments, smart health-care, energy, manufacturing, agriculture, payments, hospital, business, smart grid, and so on, as shown in Fig. 2 based on Blockchain technology with SDN-IoT architecture to enhance the Industry 4.0 applications exceedingly.

## IV. RESULTS ANALYSIS AND DISCUSSIONS

### A. Environment Setup

In this section, we have analyzed to set up the proposed model with the Emulator (Minininet), Mininet-Wifi simulation tools for measuring the environmental activities of the SDN. Again, the OpenFlow protocol is used in the SDN context to accomplish the goal outcomes. Further, different types of packet sizes are used, such as bytes 256, 800, and 1024. Nodes turn in the rectangular field according to the configuration of the random waypoint. In which, at some random location, each node is positioned the rectangular area at simulation initialization. Throughout the simulation period, all nodes shift in line with the movement set out in the scenario script. The nodes are allowed to move in the dimension 3000m x 3000m rectangular area. In addition, Ubuntu (GNU/Linux), x86 (2.20GHz), 8GB RAM, 2TB ROM, and other external memory used to test our desired performance. Importantly, the Wireshark platforms have been adequately utilized to visualize the IoT network performance based on the SDN platform.

### B. Performance Analysis

In this segment, the authors have considered three parameters, such as throughput analysis, secured rate analysis and packets failure rate comparison, to evaluate the execution of the proposed system efficiently.

*1) Throughput Analysis:* The authors have analyzed the throughput based on the number of packets transmission, as shown in Fig. 5.

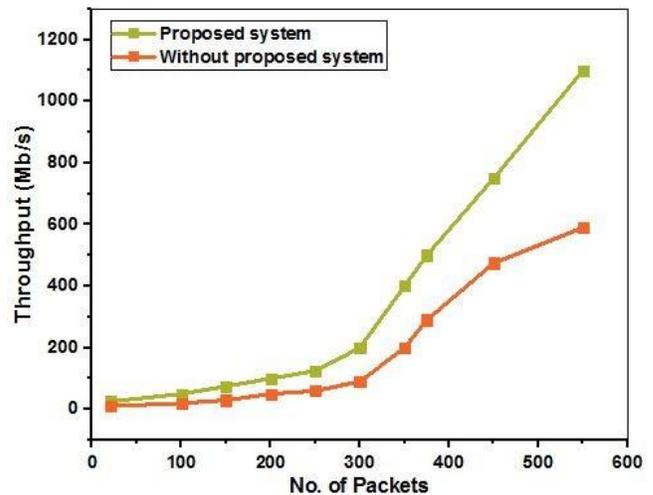

Fig. 5. Throughput Comparison

However, it displays the throughput comparisons between existing centralized model and the suggested model diagrammatically. Further, we have regarded that when the number of packets is less, then the throughput is almost the same with each other. But when the number of packets is progress- ing, then the throughput is also growing. After performing a particular time, we have also noticed that due to less engagement of attacks or undesired components like noise and intruder our proposed architecture "DistB-SDoIndustry" shows a much better performance than the centralized core model performance effectively.

*2) Security Rate Analysis:* Then, we have observed the security rate according to the number of data packets, as shown in Fig. 6. After that, it shows the comparisons of the protection rate between the core model and the proposed system "DistB-SDoIndustry". Moreover, we have also noticed that when the number of data packets is less, the security rate is nearly the equivalent. But when the number of packets increases, then both of the performance is also expanding. Furthermore, due to the rise in the number of data flow after performing a specific time, the authors have remarked that the proposed "DistB-SDoIndustry" model shows better security few involvements of attacks than the performance of the existing network.

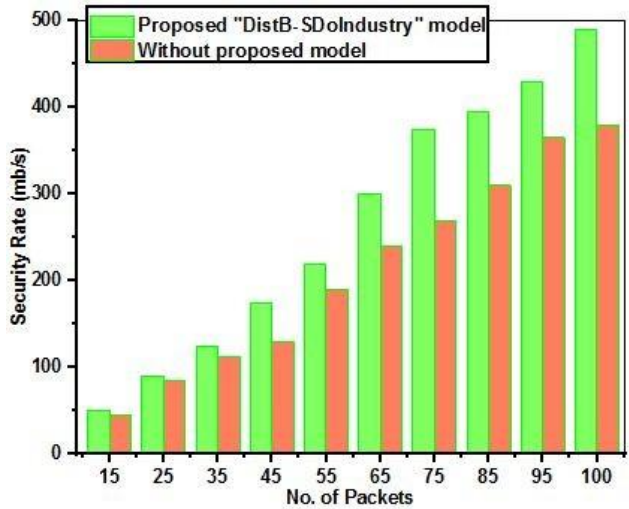

Fig. 6. Securities Comparison

*3) Node Failure Rate Analysis:* On the other hand, in Fig. 7 shows the data failure rate based on the number of nodes between our proposed system and the core system—actually, node failure rate occurs based on different types of attacks. From the above analysis, it is clear that the proposed design performs the minimal node failure rate due to less impact of attacks. Further, the authors have notified that the node failure rate is initially less for both. However, with the increase in the number of nodes, the failure rate also progresses. Besides, the authors have also observed that primarily node failure rate is 2% or 3% only for both. But after completing a few times in the core scheme's node failure rate is 90% or above, on the contrary, the proposed scheme's failure rate is 38% to 43 % only. Thus the above analysis shows that the offered secured system "DistB-SDoIndustry" overcomes failure rate significantly than the traditional core model.

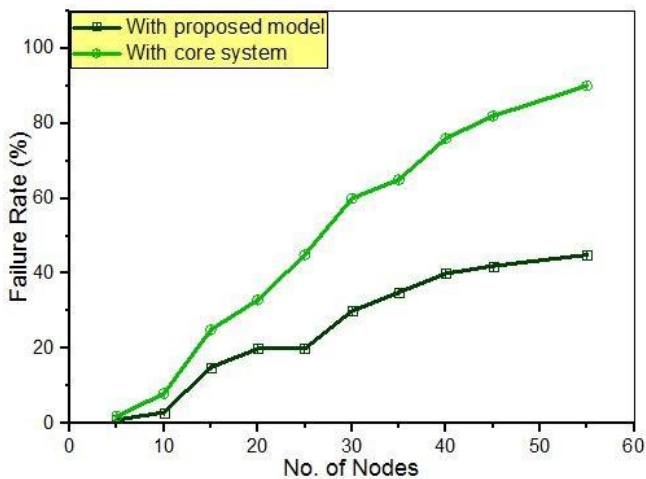

Fig. 7. Nodes Failure Comparison

## V. CONCLUSION

This paper presents a model "DistB-SDoIndustry" based on distributed Blockchain technology with SDN-IoT architecture for Industry 4.0 applications. Basically, we have highlighted two technologies such as SDN and Blockchain, in order to provide robust privacy and reliable security in the Industry 4.0 environment efficiently. Moreover, we have considered the SDN-IoT model for dividing our whole architecture into different layers securely. After that, we have also addressed Blockchain for improving data security and confidentiality. There is no interference with the third party in the presented system; furthermore, this paper has implemented the SDN-IoT model in different parameters like the security rate. This depends on no. of data packets and packets failure rate based on no. of nodes in the proposed networking model. Still, the implementation of Blockchain is in a developing stage. In the future, this study will be added to the complete implementation of Blockchain competently. Moreover, the authors will analyze the different types of attacks, like Daniel of Service (DoS) attacks, flooding attacks, etc. In addition, the authors will also evaluate more parameters such as throughput, packet arrival rate, the response time of data, etc. as well as assess the performances in numerous ways of the presented architecture.

[11] T. Han, S. R. U. Jan, Z. Tan, M. Usman, M. A. Jan, R. Khan, and Y. Xu, "A comprehensive survey of security threats and their mitigation techniques for next-generation sdn controllers," *Concurrency and Computation: Practice and Experience*, vol. 32, no. 16, p. e5300, 2020.

[12] T. Adbeb, W. Di, and M. Ibrar, "Software-defined networking (sdn) based vanet architecture: Mitigation of traffic congestion," *International Journal of Advanced Computer Science and Applications*, vol. 11, no. 3, 2020. [Online]. Available: http://dx.doi.org/10.14569/IJACSA.2020.0110388

[13] C.-S. Yang, "Maritime shipping digitalization: Blockchain-based technology applications, future improvements, and intention to use," *Transportation Research Part E: Logistics and Transportation Review*, vol. 131, pp. 108–117, 2019.

[14] R. Zhang, R. Xue, and L. Liu, "Security and privacy on blockchain," *ACM Computing Surveys (CSUR)*, vol. 52, no. 3, pp. 1–34, 2019.

[15] G. Rathee, M. Balasaraswathi, K. P. Chandran, S. D. Gupta, and C. Boopathi, "A secure iot sensors communication in industry 4.0 using blockchain technology," *JOURNAL OF AMBIENT INTELLIGENCE AND HUMANIZED COMPUTING*, 2020.

[16] Q. Wang, X. Zhu, Y. Ni, L. Gu, and H. Zhu, "Blockchain for the iot and industrial iot: A review," *Internet of Things*, vol. 10, p. 100081, 2020.

[17] T. Alladi, V. Chamola, R. M. Parizi, and K.-K. R. Choo, "Blockchain applications for industry 4.0 and industrial iot: A review," *IEEE Access*, vol. 7, pp. 176 935–176 951, 2019.

[18] A. Bagdadee, L. Zhang, and M. Remus, *A Brief Review of the IoT-Based Energy Management System in the Smart Industry*, 01 2020, pp. 443–459.

[19] K. K. Karmakar, V. Varadharajan, S. Nepal, and U. Tupakula, "Sdn enabled secure iot architecture," in *2019 IFIP/IEEE Symposium on Integrated Network and Service Management (IM)*, 2019, pp. 581–585.

[20] A. Yazdinejad, R. M. Parizi, A. Dehghantanha, Q. Zhang, and K. R. Choo, "An energy-efficient sdn controller architecture for iot networks with blockchain-based security," *IEEE Transactions on Services Computing*, vol. 13, no. 4, pp. 625–638, 2020.

[21] A. Rahman, M. K. Nasir, Z. Rahman, A. Mosavi, S. Shahab, and B. Minaei-Bidgoli, "Distblockbuilding: A distributed blockchain-based sdn-iot network for smart building management," *IEEE Access*, 2020.

[22] I. H. Abdulqadder, S. Zhou, D. Zou, I. T. Aziz, and S. M. A. Akber, "Multi-layered intrusion detection and prevention in the sdn/nfv enabled cloud of 5g networks using ai-based defense mechanisms," *Computer Networks*, p. 107364, 2020.

[23] F. H. Pohrmen, R. K. Das, and G. Saha, "Blockchain-based security aspects in heterogeneous internet-of-things networks: A survey," *Transactions on Emerging Telecommunications Technologies*, vol. 30, no. 10, p. e3741, 2019.

[24] S. B. Rane and Y. A. M. Narvel, "Re-designing the business organi- zation using disruptive innovations based on blockchain-iot integrated architecture for improving agility in future industry 4.0," *Benchmarking: An International Journal*, 2019.

[25] J. Al-Jaroodi, N. Mohamed, and I. Jawhar, "A service-oriented middleware framework for manufacturing industry 4.0," *ACM SIGBED Review*, vol. 15, no. 5, pp. 29–36, 2018.

[26] I. Jamai, L. B. Azzouz, and L. A. Saïdane, "Security issues in indus- try 4.0," in *2020 International Wireless Communications and Mobile Computing (IWCMC)*. IEEE, 2020, pp. 481–488.

[27] E. Oztemel and S. Gursev, "Literature review of industry 4.0 and related technologies," *Journal of Intelligent Manufacturing*, vol. 31, no. 1, pp. 127–182, 2020.

[28] J. Prinsloo, S. Sinha, and B. von Solms, "A review of industry 4.0 manufacturing process security risks," *Applied Sciences*, vol. 9, no. 23, p. 5105, 2019.

[29] F. Naeem, M. Tariq, and H. V. Poor, "Sdn-enabled energy-efficient routing optimization framework for industrial internet of things," *IEEE Transactions on Industrial Informatics*, 2020.

[30] A. Rahman, M. J. Islam, F. A. Sunny, and M. K. Nasir, "DistBlockSDN: A Distributed Secure Blockchain based SDN-IoT Architecture with NFV Implementation for Smart Cities," *In Press: International Conference on Innovation in Engineering and Technology (ICIET)*, vol. 23, p. 24, IEEE, 2019.

[31] R. Brozzi, D. Forti, E. Rauch, and D. T. Matt, "The advantages of industry 4.0 applications for sustainability: Results from a sample of manufacturing companies," *Sustainability*, vol. 12, no. 9, p. 3647, 2020.
682 | P a g e